\documentclass{INTERSPEECH2023}
\newcommand{\mathbbm}[1]{\text{\usefont{U}{bbm}{m}{n}#1}}
\usepackage[hang]{footmisc}
\usepackage{url}
\interspeechcameraready

\title{High-Quality Automatic Voice Over with Accurate Alignment: \\ Supervision through Self-Supervised Discrete Speech Units\thanks{\textbf{Code and voice over samples:} https://ranacm.github.io/DSU-AVO/}\thanks{This work is supported by 1) National Natural Science Foundation of China (Grant No.~62271432); 2) Guangdong Provincial Key Laboratory of Big Data Computing, The Chinese University of Hong Kong, Shenzhen (Grant No.~B10120210117-KP02); and 3) Agency for Science, Technology and Research (A*STAR) under its AME Programmatic Funding Scheme (Project No.~A18A2b0046).}} 
\name{Junchen Lu$^{1,2}$, Berrak Sisman$^2$, Mingyang Zhang$^3$, Haizhou Li$^{3,1}$}
\address{
  $^1 $National University of Singapore, Singapore \quad
  $^2 $The University of Texas at Dallas, USA \\
  $^3 $Shenzhen Research Institute of Big Data, School of Data Science,\\ The Chinese University of Hong Kong, Shenzhen, China}
\email{junchen@u.nus.edu, Berrak.Sisman@UTDallas.edu, \{zhangmingyang, haizhouli\}@cuhk.edu.cn}

\begin{document}

\maketitle

\begin{abstract}


The goal of Automatic Voice Over (AVO) is to generate speech in sync with a silent video given its text script. Recent AVO frameworks built upon text-to-speech synthesis (TTS) have shown impressive results.
However, the current AVO learning objective of acoustic feature reconstruction brings in indirect supervision for inter-modal alignment learning, thus
limiting the synchronization performance and synthetic speech quality. To this end, we propose a novel AVO method leveraging
the learning objective of self-supervised discrete speech unit prediction, which
not only provides more direct supervision for the alignment learning, but also alleviates the mismatch between the text-video context and acoustic features. Experimental results show that our proposed method achieves remarkable lip-speech synchronization and high speech quality by outperforming baselines in both objective and subjective evaluations. Code and speech samples are publicly available.

\end{abstract}
\noindent\textbf{Index Terms}: Text-to-speech, lip-speech synchronization, automatic voice over, discrete speech units, speech synthesis

\section{Introduction}

Automatic Voice Over is a cutting-edge technology that utilizes artificial intelligence to generate speech that voice-synchronizes with a pre-recorded video \cite{lu2022visualtts}. AVO technology enables the automatic creation of a voice track that is perfectly aligned with the lip movement, facial expression, and conversational tone of the video, provided with text transcription. As a highly efficient AI-powered solution for voice over, AVO has the potential to revolutionize video-making 
in various industries, including movie dubbing, online education, and marketing.

The development of neural text-to-speech synthesis has played a crucial role in the advancement of AVO technology. In light of rapid emergence of deep learning, TTS systems built upon neural networks can generate high-quality speech~\cite{tan2021survey}. 
End-to-end TTS systems, including Tacotron 1/2~\cite{wang2017tacotron, shen2018natural} and FastSpeech 1/2~\cite{ren2019fastspeech, ren2020fastspeech}, work in a simplified pipeline of mapping sequences of character or phoneme input into mel-spectrogram acoustic features. With the help of neural vocoders~\cite{oord2016wavenet, kumar2019melgan, kong2020hifi}, they generate speech with human-level naturalness.

\begin{figure}
    \centering
    \includegraphics[scale=1]{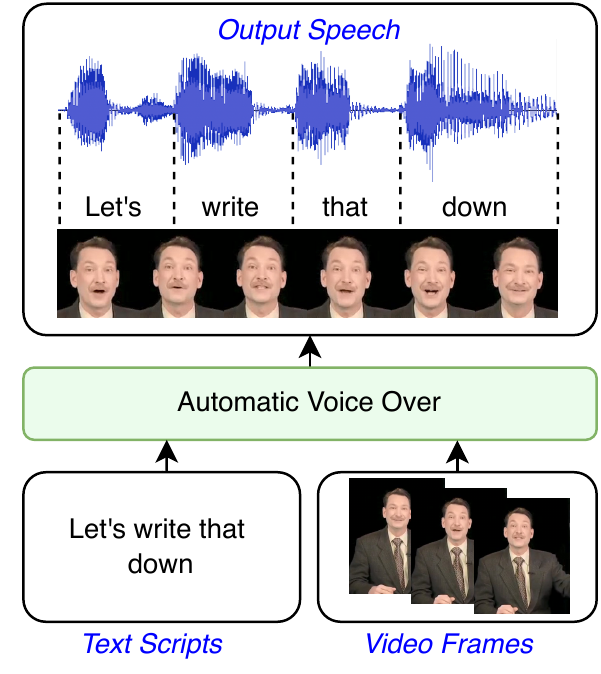}
    \vspace{-3mm}
    \caption{An illustration of the AVO workflow: The input to the system consists of video frames and corresponding text scripts, and the output is voice over speech audio in synchronization with the video.}
    \label{fig:workflow}
    \vspace{-6mm}
\end{figure}

With the ability to generate high-quality speech, neural TTS provides a foundation for AVO systems to produce accurate and natural-sounding voice over. To model speech with precise timing in sync with the video, AVO systems generally use lip motion or facial movement to guide the rendering of phonetic duration in TTS \cite{lu2022visualtts, hu2021neural, hassid2022more}, as illustrated in Figure \ref{fig:workflow}. Existing AVO approaches use attention-based alignment modules to align multi-modal features and produce text-video context for acoustic feature decoding. However, as a learning objective of an AVO system, acoustic feature reconstruction does not provide direct supervision for the model to learn accurate inter-modal alignment, 
thus affecting the acoustic decoding and leading to degradation of synthetic speech quality.

Self-supervised learning (SSL) speech models trained on large amounts of unlabeled data are proved to be sufficient for capturing content information of speech~\cite{schneider2019wav2vec, baevski2020wav2vec, hsu2021hubert, pasad2021layer}. Recent studies show that discrete units derived from SSL speech models can be applied as speech content representation for speech synthesis tasks, including speech resynthesis~\cite{polyak2021speech} and voice conversion~\cite{van2022comparison}. As speech content in AVO is modeled through the alignment of multi-modal information, content representation can provide more direct supervision for alignment learning.

Motivated by these, we propose a novel AVO approach leveraging discrete speech units: first, align multi-modal features and predict discrete speech units from the text-video context formed; then, synthesize speech conditioned on the predicted units.
The main contributions of this paper include: 
\hspace{10mm}
\begin{itemize}
    \item
    We propose a new learning objective of discrete speech unit prediction for AVO, providing more direct supervision for text-video alignment learning at the context feature level, thus improving lip-speech synchronization;
    \item
    We propose to synthesize speech directly from discrete speech units with a pretrained unit vocoder, thus alleviating mismatch between text-video context and acoustic features in acoustic decoding and improving synthetic speech quality.
\end{itemize}

\section{Related work}
\label{sec:related}

\subsection{AVO Background}

The industry of video content creation calls  for high-quality and cost-efficient voice over solutions. As such, AVO research, drawing upon TTS, has become increasingly important.
While general-purpose TTS systems generate speech from text, AVO systems need to produce speech that is not only natural-sounding but also synchronized with visual presentation of speakers in video. With this requirement, AVO systems have to model the timing of the speech, taking visual context of the video into account. 

Generally, AVO frameworks consist of text and video encoders, which encode information from multi-modal inputs, an alignment module that models inter-modal feature relationships, an acoustic decoder that constructs acoustic features such as mel-spectrogram, and a vocoder to convert acoustic features into output speech waveform ~\cite{lu2022visualtts, hu2021neural, hassid2022more}. 
Existing approaches of alignment module are based on attention mechanism \cite{vaswani2017attention}. VisualTTS \cite{lu2022visualtts} and Neural Dubber \cite{hu2021neural} align textual and visual features explicitly by attention-based aligners and form text-video context which serves as the input to acoustic decoder. VDTTS \cite{hassid2022more} uses a multi-source attention mechanism for selecting which outputs of video encoder and text encoder to pass to acoustic decoder at each decoding timestep, forming the alignment between modalities in an implicit manner. 

The existing AVO frameworks have limitations that hinder their ability to produce high-quality voice over. Firstly, the capability of the model to learn inter-modal alignment is limited to the acoustic decoding process with the learning objective of mel-spectrogram reconstruction. To establish the alignment, the entire AVO model
must be trained as a whole, which is computationally expensive and time-consuming. 
Secondly, the supervision on the acoustic feature level is indirect for alignment learning due to the gap between phonetic information presented in input text and the acoustic features to model.
Moreover, the mismatch between the context representation modeled by the alignment module and target acoustic features can negatively impact acoustic decoder training, resulting in degradation of synthetic speech quality. In this paper, we aim to address these challenges in AVO by imposing supervision on the context representation level with a learning objective of discrete speech unit prediction, which will be introduced in Sec. \ref{sec:proposed}.

\subsection{Self-supervised learning in speech synthesis}

Self-supervised learning has emerged as a powerful approach to learning speech representations. SSL speech models leverage large amounts of unlabeled speech data by defining auxiliary tasks and generating pseudo-labeled data to be trained using supervised learning techniques \cite{schneider2019wav2vec, baevski2020wav2vec, hsu2021hubert}. The pretrained models can be used for various downstream tasks~\cite{yang2022deep,yang2022KF}, such as automatic speech recognition (ASR)~\cite{pasad2021layer, chen2022fearless} and speech emotion recognition \cite{goncalves2022improving}, and speech enhancement~\cite{hsu2023revise}.

Recent studies have also explored the use of SSL speech representations in speech synthesis tasks.
Du et al. \cite{du2022vqtts} propose to reduce the complexity of the acoustic model in TTS with SSL vector-quantized acoustic features as its classification target.
Huang et al. \cite{huang2021any} propose to use discrete speech representations as a bottleneck to disentangle content information from speaker information and model acoustic features on top of them for voice conversion. 
Polyak et al. \cite{polyak2021speech} 
demonstrate that discrete units derived from HuBERT models \cite{hsu2021hubert} can serve as the input of high-quality speech waveform generation.

Inspired by these, we propose to use discrete speech units as the content representation for speech generation and the supervision of alignment learning in AVO.

\section{Proposed method}
\label{sec:proposed}
We formulate the AVO problem and propose \textit{Discrete Speech Unit-AVO (DSU-AVO)}, with motivation of providing more direct supervision to alignment learning and establishing a strong connection between text-video context and speech in AVO.

\subsection{Problem formulation}

Given input phoneme sequence $X_p = (x_{p_1}, x_{p_2}, ..., x_{p_N})$ with length $N$ and video represented by a sequence of image frames $X_v = (x_{v_1}, x_{v_2}, ..., x_{v_{T_v}})$ with length $T_v$, the goal of AVO is to generate speech audio that accurately reflects the phonetic content and is temporally aligned with the video.

In a voiced video clip, speech audio and video are both continuous signals of the same length. In practice, they are sampled at different frame rates. This allows us to encode ground-truth speech $Y$ as a sequence of speech representation $Z$ with length~$T_z$ that can be easily aligned with the video frame sequence~$X_v$ by upsampling the latter, given that the length of the speech representation sequence is $n$ times that of the video frame sequence, where $n = \frac{T_z}{T_v} \in \mathbbm{N}^+$.

Given the temporal correspondence of speech audio and video, the synchronization between speech and lip motion can be achieved by aligning phoneme information, which determines the speech content, with the lip motion information~\cite{lu2022visualtts, hu2021neural}.
An AVO framework aims to model accurate text-video alignment and produce context representation $C=f(X_p, X_v)$ with length $T_v$. Then, conditioned on the context and given the audio-video length ratio $n$, the framework generates speech representation $\hat{Z}=g(C, n)$. Finally, $\hat{Z}$ is converted to speech waveform $\hat{Y}$ through a pretrained vocoder. Training of an AVO framework can be seen as finding the optimal context modeling $f(\cdot)$ and speech representation generation $g(\cdot)$.

\subsection{Supervision at context representation level with discrete speech units}

As speech representation generation in AVO is conditioned on text-video context, context modeling $f(\cdot)$ is crucial for a framework to produce speech with accurate pronunciation and timing. Since the context is formed by aligning multi-modal representations, alignment learning is an integral part of context modeling.

Existing AVO frameworks \cite{lu2022visualtts, hu2021neural, hassid2022more} rely on acoustic features, typically mel-spectrogram, as the speech representation $Z$ to model, utilizing acoustic decoders as $g(\cdot)$.
With the learning objective of mel-spectrogram reconstruction, these frameworks guide the alignment learning through supervision at the acoustic feature level.
However, mel-spectrogram does not directly capture linguistic information \cite{shen2018natural, lee2022hierspeech}, 
posing a gap between context and speech representation, 
thus providing indirect supervision for alignment learning in AVO.

We propose to guide the context modeling and alignment learning of AVO more directly by imposing discrete speech unit prediction as the supervision at the context representation level,
given that discrete speech units are closely correlated with speech content.
Additionally, compared with mel-spectrogram, discrete speech units are more disentangled from nuisance variation \cite{polyak2021speech} and are easier to predict. With the proposed learning objective, $g(\cdot)$ essentially becomes a classification model instead of a regression model and gains better training efficiency. To be specific, we first encode inputs from different modalities, align inter-modal representations to form context, and predict discrete speech units; then, synthesize speech conditioned on the predicted units with a pretrained unit vocoder.

\subsection{DSU-AVO system}

\begin{figure}
    \centering
    \includegraphics[scale=0.84]{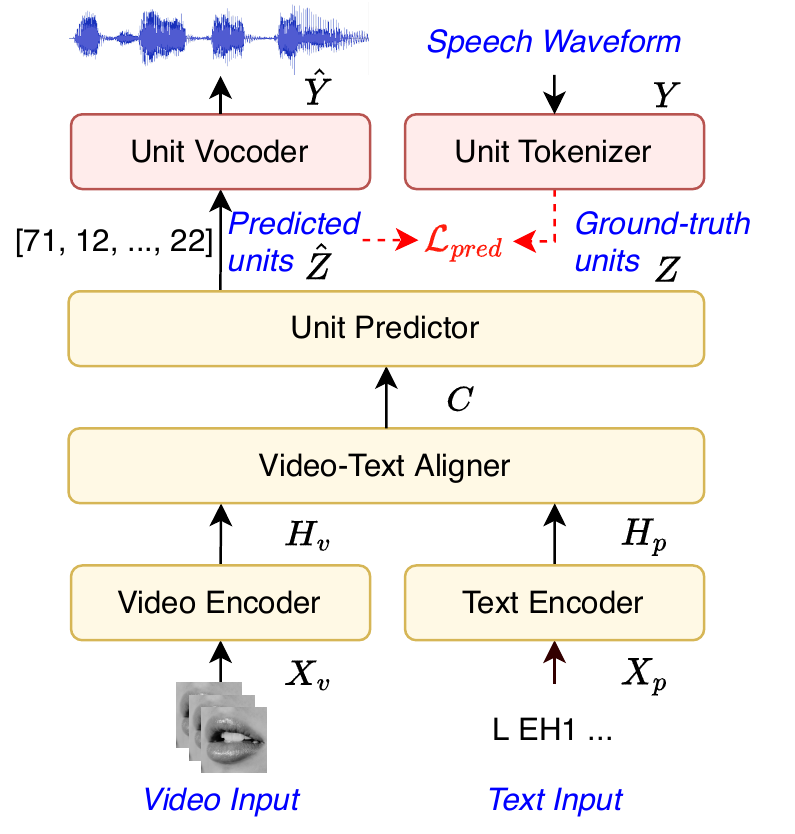}
    \vspace{-3mm}
    \caption{The model architecture of the proposed DSU-AVO. Modules denoted with red color are pretrained and frozen during DSU-AVO training. Dotted arrows denote loss calculation.}
    \label{fig:architecture}
    \vspace{-5mm}
\end{figure}

As demonstrated in Figure~\ref{fig:architecture}, our proposed DSU-AVO consists of unit tokenizer, video encoder, text encoder, video-text aligner, unit predictor, and unit vocoder.

\subsubsection{Unit tokenizer}

A HuBERT model \cite{hsu2021hubert} followed by k-means clustering is used as the unit tokenizer to encode ground-truth speech into a sequence of discrete speech units as the prediction target $Z=(z_1,z_2,...,z_{T_z})$, 
where $z_i \in \{0, 1, ...,K-1\}$ for $1 \leq i \leq T_z$ and $K$ is number of k-means centroids. The unit tokenizer is pretrained and frozen during DSU-AVO training.

\subsubsection{Encoders}

As speech progression is inherently linked with lip motion in real life \cite{lu2022visualtts, chen2021correlating}, we use lip image sequence cropped from the video clip as the input $X_v$.
The video encoder consists of a frozen visual feature extractor and feed-forward Transformer (FFT)~\cite{ren2019fastspeech, hu2021neural, vaswani2017attention} blocks. Visual features extracted by such an extractor pretrained on visual speech recognition tasks have strong correlation with phonetic information~\cite{chen2021correlating} and are efficient for aiding speech-related tasks~\cite{pan2021muse, pan2022selective}. 
Input $X_v$ is encoded into hidden visual representation $H_v \in \mathbbm{R}^{T_v \times d}$ where $d$ is the dimension of hidden representations.

We adopt the same text encoder that is used in FastSpeech~2~\cite{ren2020fastspeech} for TTS and in Neural Dubber \cite{hu2021neural} for AVO. It consists of an embedding layer followed by FFT blocks, to encode input phoneme $X_p$ into hidden textual representation $H_p \in \mathbbm{R}^{T_p \times d}$.

\subsubsection{Text-video aligner}

We utilize a text-video aligner \cite{lu2022visualtts, hu2021neural} to temporally align textual and visual representations by scaled dot-product attention \cite{vaswani2017attention}, and produce text-video context with length $T_v$:
\begin{subequations}
\vspace{-2mm}
\begin{align}
    C=&{\rm softmax}(\frac{H_{v}H_{p}^T}{\sqrt{d}})H_{p}+H_v \\= & AH_{p}+H_v \in \mathbbm{R}^{T_v \times d}
\end{align}
\end{subequations}
where $H_{v}$ serves as the query, $H_{p}$ serves as the key and the value, and $A \in \mathbbm{R}^{T_v \times T_p}$ is the attention weight matrix. $H_v$ is added through residual connection to enhance alignment learning. $C$ is then upsampled to match $T_z$ by simply duplicating each frame of representation $n$ times, where $n$ is the audio-video length ratio.

We adopt the diagonal constraint loss $\mathcal{L}_{diag}$ following \cite{hu2021neural, chen2020multispeech} to shape diagonal attention.
\vspace{-2mm}
\subsubsection{Unit predictor}

The accuracy of the prediction has a direct impact on the content correctness and intelligibility of the synthetic speech. A higher prediction accuracy results in a better perceptual quality of the synthetic speech. Hence, DSU-AVO focuses on predicting accurate discrete speech units. 

The unit predictor consists of an FFT block
that further models the context into a more deterministic representation for accurate unit prediction, 
followed by a softmax layer that maps output of the FFT block onto a probability distribution over a set of output classes, i.e., the discrete speech units $\{0,1,...,K-1\}$. 
Taking the context $C$ produced by the text-video aligner as input, this module predicts a sequence of discrete speech units $\hat{Z}=(\hat{z_1}, \hat{z_2}, ..., \hat{z}_{T_z})$ with the same length of ground-truth discrete speech units $Z$.
The prediction is guided by minimizing the cross entropy loss $\mathcal{L}_{pred}$: 
\vspace{-2mm}
\begin{equation}
    \mathcal{L}_{pred} = -\sum\nolimits_{t=1}^{T_z} \sum\nolimits_{k=0}^{K-1} \mathbf{z}_{t,k} {\rm log} p_{t,k} (C)
    \vspace{-1mm}
\end{equation}
where $\mathbf{z}_t$ is the one-hot vector representing speech unit truth label at the $t$-th frame, and $p_{t,k}(C)$ is the predicted softmax probability for unit $k$ at the $t$-th frame, given context $C$.
Finally, the overall training criterion is $\mathcal{L} = \mathcal{L}_{pred} + \mathcal{L}_{diag}$.

\subsubsection{Unit vocoder}
We utilize Unit HiFi-GAN \cite{polyak2021speech}, which is pretrained on ground-truth $<$units, waveform$>$ pairs without the speaker encoder and the F0 encoder in a single-speaker setting~\cite{lee2022direct}, as the unit vocoder.
Given predicted units $\hat{Z}$, it generates speech audio as the output $\hat{Y}$ of AVO. As Unit HiFi-GAN models the mapping from discrete speech units to speech waveform without any spectrogram estimation~\cite{polyak2021speech}, DSU-AVO does not need to model acoustic features for speech generation, thus alleviating the mismatch between context and acoustic features.

\section{Experiments}
\label{sec:experiments}

\subsection{Experimental setup}

\subsubsection{Dataset and preprocessing}

We utilize Chem dataset \cite{hu2021neural} from Lip2Wav \cite{prajwal2020learning} to evaluate the performance of AVO frameworks. Chem dataset is a single-speaker audio-visual English speech dataset with official transcripts collected from YouTube. We use 6088 samples for training, 200 samples for validation, and 200 samples for testing. As a preprocessing step, we follow the same process described in~\cite{shilearning} to crop 88 $\times$ 88 lip region-of-interest for the video input. All video clips are sampled at 25Hz frame rate, and audio samples are sampled at 16kHz.

\subsubsection{Model configurations}

We evaluate several systems, including: 
1) Mel Resynthesis, where we convert the ground-truth audio into mel-spectrogram and convert it back to waveform with HiFi-GAN\footnote{https://github.com/jik876/hifi-gan} \cite{kong2020hifi}; 
2) Unit Resynthesis, where we synthesize speech audio conditioned on ground-truth discrete speech units using Unit HiFi-GAN\footnote{https://github.com/facebookresearch/speech-resynthesis} \cite{polyak2021speech}; 
3) FastSpeech 2 \cite{ren2020fastspeech}, a TTS baseline\footnote{\label{link:fs2}https://github.com/ming024/FastSpeech2} that generates speech conditioned only on text, without taking visual information into consideration; 
4) Neural Dubber \cite{hu2021neural}, an AVO baseline with the learning objective of mel-spectrogram reconstruction; and 5) DSU-AVO, our proposed framework. As there is no publicly available implementation of Neural Dubber, we implement the framework based on an open-source FastSpeech~2 implementation\textsuperscript{\ref{link:fs2}} without the image-based speaker embedding module as we conduct AVO in a single-speaker setting. 

System 1), 3), and 4) use the same HiFi-GAN for a fair comparison. 
System 2) and 5) use the same Unit HiFi-GAN.
Both vocoders are trained on Chem dataset. 
We set the number of FFT blocks in text encoders of 3), 4), and 5) to 4, the one in video encoders of 4) and 5) to 2, the one in decoders of 3) and 4) to 6.
$d$ is set to 256 for 3), 4), and 5). For visual feature extractor in both 4) and 5), we use the same AV-HuBERT + Self-Training model\footnote{https://github.com/facebookresearch/av\_hubert}~\cite{shilearning} pretrained on 1,758h of unlabeled Voxceleb2 data~\cite{chung2018voxceleb2} and finetuned on 433h of labeled LRS3 data~\cite{afouras2018lrs3} for visual speech recognition. The extracted visual feature is projected to $d$ dimensions as the input to FFT blocks in video encoder. We note that in our implementation, Neural Dubber and DSU-AVO use identical text encoder, video encoder, and text-video aligner for a fair comparison. 
In DSU-AVO, we use a HuBERT Base model \cite{hsu2021hubert} pretrained on 960h LibriSpeech corpus \cite{panayotov2015librispeech} and an accompanying k-means model trained on LibriSpeech clean-100h dataset \cite{panayotov2015librispeech} as the unit tokenizer, following \cite{polyak2021speech}. Ground-truth speech is encoded to discrete speech units with $K=100$ centroids at 50Hz.

\subsection{Experimental results}

\begin{table}
    \caption{Evaluation results of LSE-C, LSE-D, FD, WER, and MOS (with 95\% confidence intervals). Arrows indicate whether higher or lower metric values are better.}
    \centering
    \resizebox{0.47\textwidth}{0.16\linewidth}{
    \begin{tabular}{c|c @{\hspace{0.5em}} c @{\hspace{0.5em}} c @{\hspace{0.5em}} c @{\hspace{0.5em}} c}
    \hline
     Method & LSE-C $\uparrow$ & LSE-D $\downarrow$ & FD $\downarrow$ & WER(\%) $\downarrow$ &  MOS $\uparrow$                      \\ \hline \hline
     Ground Truth   & 7.00 & 7.31 & NA & 11.4 & 4.69 $\pm$ 0.06 \\ 
     Mel Resynthesis & 6.89 & 7.40 & 0.66 & 11.8 & 4.58 $\pm$ 0.07 \\ 
     Unit Resynthesis & 6.99 & 7.39 & 0.82 & 20.6 & 4.06 $\pm$ 0.08 \\\hline
     FastSpeech 2 \cite{ren2020fastspeech}             &  2.69   &  11.78    &  40.38 & 25.4 & 3.10 $\pm$ 0.09 \\
     Neural Dubber \cite{hu2021neural}             &  6.11  &  8.47  & 9.39  & 75.8 & 2.43 $\pm$ 0.12  \\
     \textbf{DSU-AVO}                    &   \textbf{6.81}   &  \textbf{7.56} & \textbf{3.23}  &  \textbf{24.7}  & \textbf{3.98 $\pm$ 0.08}\\
    
    \hline
    \end{tabular}}
    \label{tab:eval}
    \vspace{-3mm}
\end{table}

\begin{table}
    \caption{Evaluation results of the BWS listening test on lip-speech synchronization. N/P stands for no preference.}
    \centering
    \resizebox{0.36\textwidth}{0.10\linewidth}{
    \begin{tabular}{c|c c c}
    \hline
     Method & Best(\%) & Worst(\%) & N/P(\%) \\ \hline \hline
     FastSpeech 2 \cite{ren2020fastspeech}  &  4.0 &  72.0 & 24.0           \\
     Neural Dubber \cite{hu2021neural}  &  12.0  &  26.7  &  61.3  \\
     \textbf{DSU-AVO} & \textbf{84.0} &  \textbf{1.3} & 14.7 \\
    
    \hline
    \end{tabular}}
    \label{tab:bws}
    \vspace{-4mm}
\end{table}

\subsubsection{Objective evaluation}

We measure lip-speech synchronization between the synthetic speech and input video with Lip Sync Error - Confidence (LSE-C) and Lip Sync Error - Distance (LSE-D) \cite{prajwal2020lip}, using a pretrained SyncNet model \cite{chung2017out}. LSE-C denotes the confidence score of audio-video synchronization time offset, where higher values indicate more accurate synchronization. LSE-D measures the distance between audio and video features, where lower values indicate better synchronization. As shown in Table~\ref{tab:eval}, DSU-AVO outperforms the baselines by achieving LSE-C of 6.81, and LSE-D of 7.56.

We utilize Frame Distrubance (FD) \cite{9262021} to measure the duration deviation between generated speech and ground-truth speech from the test set. Since ground-truth speech is in sync with video, FD also indicates lip-speech synchronization performance for AVO \cite{lu2022visualtts}. We note that DSU-AVO exhibits remarkable results and outperforms baselines with an FD of 3.23. LSE-C, LSE-D, and FD results prove the effectiveness of DSU-AVO in alignment learning.

We report the Word Error Rate (WER) obtained by the Wav2Vec 2.0 Large ASR model\footnote{https://github.com/facebookresearch/fairseq/tree/main/examples} \cite{baevski2020wav2vec} pretrained and finetuned on 960h LibriSpeech \cite{panayotov2015librispeech} data as an assessment of synthetic speech intelligibility. Note that WER is for relative comparison only, since the ASR model is not finetuned on Chem dataset. DSU-AVO achieves a WER value of 24.7, demonstrating a strong ability to model correct speech content given input text.

\subsubsection{Subjective evaluation}

Human perception plays a crucial role in evaluating the performance of AVO, as the goal of AVO is to generate speech that appears natural to human observers. We conduct listening experiments, in which 15 listeners participate. In the mean opinion score (MOS) evaluation, each participant listens to 12 speech samples produced by each system and rate the speech audio quality on a five-point scale. As shown in Table \ref{tab:eval}, our proposed DSU-AVO produces speech with a higher level of naturalness than both baselines by achieving a MOS score of 3.98~$\pm$~0.08.

We also conduct a Best-Worst Scaling (BWS) test \cite{9262021} on lip-speech synchronization, where each subject watches in total 10 scaling sets of videos and selects the best and the worst lip-speech synchronization from each set. The original pre-recorded speech samples in the test set videos are replaced with synthetic speech samples produced by system 3), 4), and 5).
Table \ref{tab:bws} shows that DSU-AVO outperforms baselines in terms of lip-speech synchronization, with the highest best votes (84.0\%) and the lowest worst votes (1.3\%).

\section{Conclusions}
\label{sec:conclusions}

In this paper, we propose DSU-AVO, a novel AVO approach leveraging discrete speech units as the content representation. Our proposed method not only provides more direct supervision for alignment learning, but also alleviates the mismatch between context and acoustic features. Experimental results show that DSU-AVO outperforms baselines in terms of synthetic speech quality and lip-speech synchronization. In future work, we will investigate further modeling speech expressiveness with SSL speech representations for AVO.

\bibliographystyle{IEEEtran}

\end{document}